# Allometric scaling law and ergodicity breaking in the vascular system

**Michael Nosonovsky[1,2]\* and Prosun Roy[1]**

[1]Department of Mechanical Engineering, University of Wisconsin-Milwaukee,
3200 N Cramer St, Milwaukee WI 53211, USA

[2]X-BIO Institute, University of Tyumen, 6 Volodarskogo St, Tyumen, 625003, Russia

**\*** Corresponding author: nosonovs@uwm.edu; Tel.: +1-414-229-2816

**Abstract**

Allometry or the quantitative study of the relationship of body size to living organism physiology is an important area of biophysical scaling research. The West-Brown-Enquist (WBE) model of fractal branching in a vascular network explains the empirical allometric Kleiber law (the ¾ scaling exponent for metabolic rates as a function of animal's mass). The WBE model raises a number of new questions, such as how to account for capillary phenomena more accurately and what are more realistic dependencies for blood flow velocity on the size of a capillary. We suggest a generalized formulation of the branching model and investigate the ergodicity in the fractal vascular system. In general, the fluid flow in such a system is not ergodic, and ergodicity breaking is attributed to the fractal structure of the network. Consequently, the fractal branching may be viewed as a source of ergodicity breaking in biophysical systems, in addition to such mechanisms as aging and macromolecular crowding. Accounting for non-ergodicity is important for a wide range of biomedical applications where long observations of time series are impractical. The relevance to microfluidics applications is also discussed.

**Keywords**: allometry, ergodicity, fractal branching, capillaries, cardiovascular system, microfluidics

## 1. Introduction

Modern high throughput biomedical and bioinformatics methods often require *in vivo* observations of temporal evolution of various parameters. In many cases, prolonged data acquisition from an object is impractical and it is substituted by taking a snapshot from many similar objects. For this reason, understanding the ergodicity breaking mechanisms is particularly important for the study of biophysical fluid transport systems.

Ergodicity, or the equivalence of time and phase space (or ensemble) averages, is an important property of many dynamical systems. Roughly speaking, ergodic systems have no memory of their previous history, and they tend to attain all microstates available. Contrary to that, non-ergodic systems demonstrate evolution with time or aging, which affects their ability to attain microstates with equal probability. The concept of ergodicity was first introduced by Ludwig Boltzmann, and it was further advanced in 1890 by Henri Poincaré's Recurrence Theorem stating that phase-space volume-preserving systems will always return to a state identical or very close to their initial state.

George Birkhoff (1931) and von Neumann further extended the Ergodic theory, which became a central part of the theory of dynamical systems. Ergodicity is related to a number of important concepts in $20^{th}$ century physics, such as the theory of spontaneous symmetry breaking and phase transitions. For example, it has been realized that a finite system returning to its initial position over a sufficiently long period of time makes phase transitions prohibitive in finite systems represented by Ising-type models (Kadanoff 2009).

Ergodicity has various implications for the theoretical aspects of dynamical system's behavior, such as its stability and quasi-periodic motion, and for its qualitative behavior (Arnold 1968, 1978), e.g., for identifying Lagrangian Coherent Structures (LCSs) in fluid flow (Rypina et al. 2011). Ergodicity is also crucial for practical aspects of measuring systems parameters, since sufficiently long observations of temporal behavior is often impossible and, therefore, it should be substituted with finite time measurements (Guzman-Sepulveda et al. 2017; Magdziarz and Zorawik 2019).

One area where ergodicity breaking is particularly important is biophysical transport of liquids. This includes hemodynamics (blood flow dynamics), intracellular and extracellular transport of complex media in biological systems, such as cytoplasm and nucleoplasm, and macromolecular biopolymer solutions (Kulkarni et al. 2003; Földes-Papp and Baumann 2011; Manzo et al. 2015).

From the thermodynamic point of view, biological systems including the vascular system may be viewed as open systems, so ergodicity is not expected in them, although the assumption of ergodic behavior is often made. Ergodicity breaking in biological fluids is associated with macromolecular crowding, which causes the anomalous diffusion. According to the classical Einstein – von Smoluchowski model, the mean-square displacement is a linear function of the lag time, $\langle r^2 \rangle \propto t$. Contrary to that, the anomalous diffusion results in the dependency which has the form of a power law

$$\langle r^2 \rangle \propto t^\alpha \tag{1}$$

with α<1 for the subdiffusion (which is the common case) (Hofling and Franosch 2013).

The subdiffusion caused by aging can have several underlying causes besides the macromolecular crowding. This includes flowing through obstacles with a certain density. Another cause can be the so-called "hydrodynamic memory" when a particle's effective mass should be adjusted due to the deceleration caused by incessantly new vortices diffusing slowly through the fluid. Consequently, the friction force depends on the entire history of the particle's trajectory, and it is related to the fractal nature of a turbulent trajectory (Hofling and Franosch 2013). In all these cases, the fractal behavior is responsible for ergodicity breaking and for transport deceleration in comparison with the classical diffusion law.

Besides the turbulence and random-walk diffusive behavior both leading to the fractal geometry of the trajectories, the self-similar or self-affine behavior could be associated with another situation, which is rarely viewed as a source of ergodicity breaking. This is the self-similar tree-shaped flow in the cardiovascular or alveolar systems (Bejan 2004), which is believed to be responsible for the allometric scaling in living organisms.

The allometric scaling relationships were summarized by the empirical Kleiber law. Kleiber (1932, 1947) compared metabolism rates, *B*, in various species and found that it is well approximated by a power-law scaling dependency on the mass of an animal, $B \propto M^{0.75}$. The value of the exponent, $a=0.75$, could not be explained until the seminal paper by West et al. (1997) appeared, which used a fractal model of the branching of blood vessels.

The theory by West et al. (1997), also referred to as the West-Brown-Enquist (WBE) model, caused some discussions in literature (Savage et al. 2004; Kozlowski and Konarzewski 2004; Bejan 2004; Brown et al. 2005; Etienne et al. 2006, 2008; Banavar et al. 2010); however, despite its shortcomings, the WBE model remains the main explanation of the allometric scaling exponents. At the same time, the WBE model has never been viewed as an underlying reason for ergodicity breaking in hemodynamics. While dealing with capillaries, the WBE model does not take into account the physico-chemical capillary phenomena, such as the effect of the surface tension, associated with the flow of such a multi-component liquid as blood. Instead, the WBE model relies solely on macroscale fluid mechanics considerations. In the present paper we extend the allometric model to include the capillary phenomena and investigate ergodicity breaking caused by the fractal nature of the capillary branching model.

**2. Scaling relationships for a branching capillary system**

Scaling relationships traditionally play a significant role in biophysics, since scaling considerations define the size and properties of living cells (Fabry et al. 2003; Bormashenko and Voronel 2018). A particularly important area of biophysical scaling is the allometry or the quantitative study of the relationship of body size to living organism physiology.

*2.1. Branching with area and volume conservation*

West et al. (1997) suggested an allometric scaling law for a branching cardiovascular network based on the assumptions of area- and volume-preserving branching (the WBE model). The concept of the area-preserving branching has been known in biology for a long time. Already Leonardo da Vinci suggested that in trees, the total cross section area of branches is conserved across branching nodes. This structure is believed to be self-similar and optimized to resist wind-induced loads (Eloy 2011). The volume preservation implies that the same volume is served equally by branches of different sizes.

According to the WBE model, when a tube with the length $l_k$ and radius $r_k$ branches into $n$ tubes with the lengths $l_{k+1}= \gamma l_k$ and radii $r_{k+1}= \beta r_k$, the volume served by the next generation tubes and their cross-section area should be conserved, which leads to the scaling relationships $\gamma \propto n^{-1/3}$ and $\beta \propto n^{-1/2}$. The volume is preserved because the same volume in the organism is served by blood vessels of different hierarchical levels. The area is preserved on the assumption of the constant rate of the fluid flow at different hierarchical levels (**Fig. 1**).

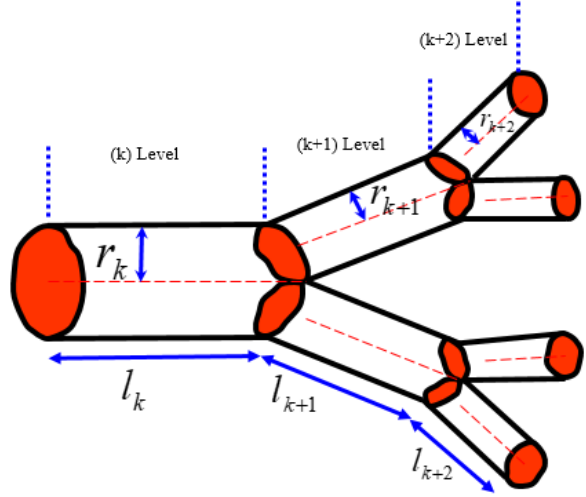

Figure 1. Branching of a vascular network; three levels are shown.

The volume of fluid at a certain *k*-th level of the branching hierarchy $V_k = \pi r_k^2 l_k n^k = V_0(n\gamma\beta^2)^k$, where $V_0 = \pi r_0^2 l_0$. The total volume of fluid is given by the sum of the geometric series $V = \sum_{k=0}^{N} \pi r_k^2 l_k n^k = V_0 \frac{1-(n\gamma\beta^2)^{N+1}}{1-n\gamma\beta^2}$, where *N* is the total number of branch generations. Since $n\gamma\beta^2 < 1$ and $N \gg 1$, a good approximation is $V = V_0 \frac{1}{1-n\gamma\beta^2} = V_c \frac{(\gamma\beta^2)^{-N}}{1-n\gamma\beta^2}$ where $V_c = V_0/(\gamma\beta^2)^{-N}$ is a certain elementary volume (e.g., volume served by a capillary) (West et al. 1997). Therefore, the volume scales as $V \propto (\gamma\beta^2)^{-N}$. From this, the scaling dependency of the total number of capillaries as a function of volume is $n^N \propto V^a \propto (\gamma\beta^2)^{-Na} \propto (n^{-4/3})^{-Na} \propto n^{4Na/3}$ yielding *a*=3/4, the well-established empirical results known as the Kleiber law.

The total volume of fluid in the cardiovascular network was further assumed to be linearly proportional to the mass of the organism. At the same time, the number of capillaries is proportional to the flow rate and to the rate of metabolic processes in general. This leads to various conclusions, for example, that the lifespan scales with the mass of an animal as $M^{1/4}$ (Bejan 2012).

*2.2. Considering more realistic scaling dependencies*

Note that the WBE branching model takes into account neither the physico-chemical capillary effects due to the interfacial tension of the liquids, nor the tortuosity of the capillaries. Human blood is a non-Newtonian multi-component fluid consisting of liquid plasma (55% of total blood volume, consisting of water by 95% with various dissolved substances), red blood cells or erythrocytes (flexible disks of 6-8 μm diameter), white blood cells or leucocytes of different types (of 7-30 μm diameter), and platelets (of 2-3 μm diameter). However, the WBE model does not properly consider the effect of such a complex flow. Moreover, the radius of blood vessels The radius of blood vessels varies by about 3000 times from 15 mm in the aorta to 5 μm in the capillaries, while the flow velocity changes by about 1300 times from 0.4 m/s in the aorta to 0.3 mm/s in the capillaries. This is in striking contradiction with the assumption of the constant rate of the fluid flow at different hierarchical levels. Therefore, more accurate scaling dependencies than those of the WBE model should be studied.

Computation Fluid Dynamics (CFD) simulations demonstrate that even simple branching (bifurcation, *n*=2) of a vessel results in a complex flow velocity profile. Thus, **Fig. 2** shows CFD modeling results for bifurcation of a blood vessel for typical blood flow conditions at the peak systolic phase. The velocity profile and flow streamlines are far from being uniform. Therefore, the scaling assumption for velocities in a branching network can be satisfied only approximately.

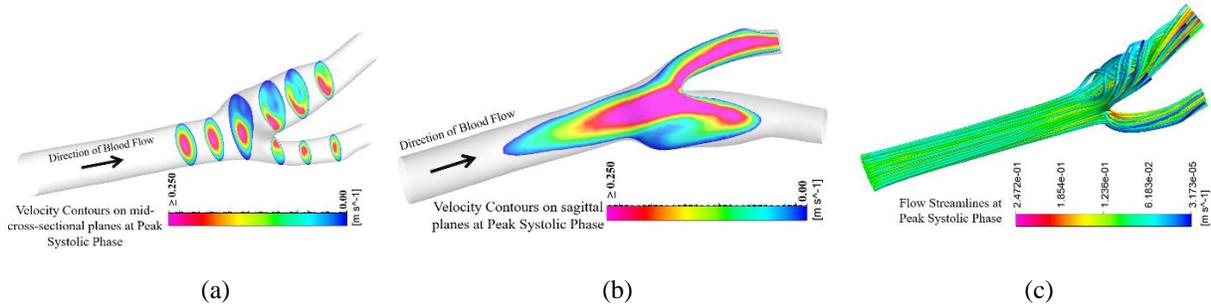

(a) (b) (c)

Figure 2. CFD simulation of blood flow through a bifurcating (branching) vessel shows complex dependencies of velocity, which cannot be approximated by a simple scaling relationship. Velocity contours in (a) a perpendicular and (b) a parallel cross-section along with (c) flow streamlines for the peak systolic phase are shown.

Due to the liquid volume preservation postulate, $\pi r_k^2 u_k = n\pi r_{k+1}^2 u_{k+1}$, the velocity scales as $\frac{u_k}{u_{k+1}} = n\frac{r_{k+1}^2}{r_k^2} = n\beta^2$. Under the assumption of cross-sectional area preservation, $\beta \propto$ $n^{-1/2}$, this leads to a constant value of the flow velocity in the system, $u_k = u$; however, experimental data indicate that the flow velocity may be different at different branching levels, as was discussed in the preceding section. As suggested by the CFD simulations and experimental data, the decrease of the flow speed is roughly of the same order as vessel radius and can be captured with the scaling assumption $\beta \propto n^{-1/3}$.

The surface tension of blood is rarely discussed in biomedical literature, although it is significantly lower than that of water: about 56 mN/m at room temperature, while water surface tension is about 72 mN/m. The surface tension of blood has biological and medical implications, thus, it has been associated with the genesis of the decompression sickness and other processes in the human organism (Hrnčíř and Rosina 1997; Krishnan et al. 2005). Therefore, it is important to pay attention to blood surface tension.

The superhydrophobicity, or the surface roughness-induced non-wetting, has been studied intensively in two recent decades (Nosonovsky and Rohatgi 2012). This brought attention to related ideas of the surface roughness-induced self-cleaning in the liquid flow, the oleophobicity (repelling organic liquids such as oils), underwater oleophobicity to reduce fouling, and the shark skin effect (flow drag reduction due to specially oriented micro-riblets). Maani et al. (2015) suggested the micro/nanostructure-controlled adhesion in blood flow for cardiovascular applications, where it is desirable to reduce stagnation and clotting of blood. The complex structure of blood vessel surface layers combined with a complex multiphase composition of blood may result in significant surface-induced effects (**Fig. 3**) Biomedical "hemophobic" applications can prevent blood clotting and thrombosis by controlling the surface pattern at a wall of a catheter or stent (Ramachandran et al. 2015). Taking into account the surface effect may modify the law of cross-sectional area preservation for small capillaries, leading to the alternate assumption that $\beta \propto n^{-1}$.

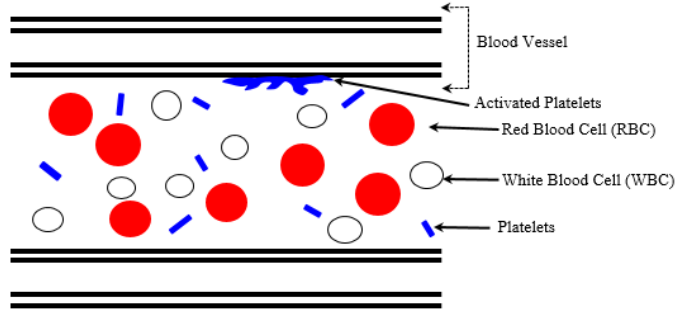

Figure 3. Schematic of blood flow in a blood vessel showing layers of the blood vessel and various components of blood. The interaction with the blood vessel wall is important due to its medical significance and hemodynamic effects (Maani et al. 2015).

Times spent by a particle or molecule at different levels of branching hierarchy form a geometric sequence with the ratio

$$\frac{T_{k+1}}{T_k} = \frac{l_{k+1}u_k}{u_{k+1}l_k} = \gamma n\beta^2 \qquad (2)$$

To accommodate for the vascular resistance, the Hagen–Poiseuille equation is often used, which states that the resistance to the blood flow scales as $r^{-4}$, or the Thurston (1976) equation, which scales the flow resistance as $r^{-3}$. There are different ways to account for the size effect due to the capillary phenomena; however, in general the assumption of the dependencies of γ and β on *n* should be modified.

In addition, capillaries are not straight tubes. The influence of capillary tortuosity or curvature can be scale-dependent. To account for various above-mentioned effects and the corresponding various possible values for the scaling rates, we can consider more general scaling relationships $\gamma \propto n^{-b}$ and $\beta \propto n^{-c}$. In other words, since different assumptions lead to different relationships between γ, β and *n*, we will now consider a general case of the relationships. Repeating the same derivation as in the preceding section, one now obtains $n^N \propto V^a \propto (\gamma\beta^2)^{-Na} \propto (n^{-b-2c})^{-Na} \propto n^{Na(b+2c)}$ leading to the power exponent

$$a = 1/(b + 2c) \qquad (3)$$

The relationship given by Eq. 3 is based on the fractal geometry of branching vessels, and in the case of *b*=1/3, *c*=1/2 it produces the same result as the WBE model, *a*=3/4.

**3. Ergodicity breaking analysis**

Let us now concentrate on the ergodicity of the blood flow in the vascular system. For that end, we will calculate the fraction of time that a blood cell spends in a certain fraction of the total volume of the circularly system, and, after that, we will compare the temporal and volumetric fractions.

The total time that a blood particle (including cells or molecules) spends in the circulatory system is given by the sum of the infinite geometric series with the ratio obtained from Eq. 2 as $T_{k+1}/T_k = \gamma n \beta^2 = n^{1-b-2c}$ The sum is given by (Wolfram 2020)

$$T = \sum_{k=1}^{\infty} T_k = \sum_{k=1}^{\infty} \frac{l_k}{u_k} = \frac{l_1}{u_1(1-n^{1-b-2c})} \tag{4}$$

where $l_1$ is the length of the highest hierarchical level. For $b=1/3$ and $c=1/2$, Eq. 4 yields $T = \frac{l_1}{u(1-1/\sqrt[3]{n})}$.

Let us calculate the time, which a blood particle spends at the levels of the system starting from $k = m$ to $k = \infty$ using the formula of a partial sum of a geometric series. The assumption behind this calculation is that only blood vessels $k = m, \ldots, \infty$ serve the volume defined by the length $l_k$ (the volume itself is given by the sphere $32\pi l_m^3/3$). Consequently, the ratio of the time spent at the levels $k = m, \ldots, \infty$ to the total time in the vascular system, $T$, is given by the time fraction

$$\theta_m = \left(\frac{T_{k+1}}{T_k}\right)^{m-1} = n^{(1-b-2c)(m-1)} \tag{5}$$

Note that for $b=1/3$ and $c=1/2$, Eq. 5 yields $\theta_m = n^{-(m-1)/3}$.

Consider a region with the radius $= 2l_m$, which corresponds to the volume $32\pi l_m^3/3$ in the organism, served by capillaries of the levels $k \geq m$. This volume constitutes the fraction of the total volume

$$\rho_m = \frac{l_m^3}{l_1^3} = \gamma^{3(m-1)} = n^{-3b(m-1)} \tag{6}$$

The volume distribution of the capillaries is uniform, therefore, the probability to find a liquid molecule or cell/particle in a given volume is supplied by $\rho_m$. On the other hand, the time fraction spent by a single blood cell or molecule at a given volume $32\pi l_m^3/3$ is given by $\theta_m$, and thus the ratio of the probability to find a single cell or molecule within a certain volume during its motion to the fraction of molecules in this volume is

$$\frac{\theta_m}{\rho_m} = \frac{n^{(1-b-2c)(m-1)}}{n^{-3b(m-1)}} = n^{(1+2b-2c)(m-1)} \tag{7}$$

In general, the ratio supplied by Eq. 7 is dependent on $m$, and, therefore, it is dependent on the size of the volume thus indicating that the motion is not ergodic. Due to the fractal nature of the capillary system, a particle tends to spend much more time in smaller volumes in comparison to the volume fraction. Note that for $b=1/3$ and $c=1/2$, Eq. 7 yields

$$\frac{\theta_m}{\rho_m} = n^{2(m-1)/3} \tag{8}$$

Using $3b(m-1)\ln n = -\ln\rho_m$ we find $m-1 = -\ln\rho_m/(3b\ln n)$. By further substituting the value of $m$ into Eq. 7 and finding a logarithm, we obtain

$$\ln\theta_m = \frac{b-1+2c}{3b}\ln\rho_m + \ln(n^{-1+b+2c}) \tag{9}$$

The number of the hierarchical level $m$ is eliminated from Eq 9, which presents the dependency of the time fraction spent in a certain volume upon its volume fraction of the total volume. By interpolating it (i.e., assuming continuous rather than discrete ranges of $\theta$ and $\rho$), one finds the dependency of the probability for a particle to be found in a certain volume based on the time spent there to the probability to find a particle in that volume based on the volume fraction

$$\theta = \rho^{\frac{b-1+2c}{3b}} \tag{10}$$

Ergodicity implies that the ensemble average equals the time average. For blood particles (such as blood cells or molecules), this requires a comparison of how many particles (as a fraction of the total number) are found in a certain volume vs. how much time (as a fraction of the total time) one particle spends in the volume. For a given volume, the space-average probability to find a particle is proportional to that volume. Note that $\rho$ is the fractional volume of the tissue, which is served by capillaries. When experimental observations are performed to trace a particle, the parameter of interest is the volume of the tissue. The volume of liquid in the vascular circulatory system, which serves the volume of the tissue, is a different volume. The capillaries in our model (similarly to the WBE model) provide equal access to all tissues. Consequently, the ensemble-averaged probability to find a blood particle in a certain volume of tissue, $\rho$, is just proportional to the volume. Contrary to that, the time fraction, $\theta$, is not necessarily proportional to the volume of the tissue. This is because a particle spends more time in smaller regions due to the self-similar scaling of a fractal branching network.

Three dependencies of the time fraction spent in a certain volume upon its volume fraction, $\theta(\rho)$, are shown in **Fig. 4(a)** for different values of *a* and *b*. Note that for *b*=1/3 and *c*=1/2, Eq. 10 yields a non-linear (and, therefore, non-ergodic) dependency $\theta = \sqrt[3]{\rho}$.

The ergodicity condition would imply equal time spent in equal volumes and, therefore, a linear dependency between corresponding variables, $\theta \sim \rho$. This is achieved when the following condition is satisfied

$$b = c - \frac{1}{2} \qquad (11)$$

For example, in **Fig. 4(a)**, the linear dependency for $b = 1/4, c = 3/4$ is the ergodic case. Combining Eqs. 3 and 11 gives $a = 2/(6c - 1)$. For volume-preserving branching, *b*=1/3, this yields $c = 5/6$ and $a = 1/2$. This is inconsistent with the Kleiber law, which relates the metabolic rates to the animal mass, $B \propto M^a$ as *a*=3/4. On the other hand, the Kleiber law, *a*=3/4, is satisfied when $b = 25/22 \approx 1.136$ and $c = 18/11 \approx 1.636$, which contradicts the volume-preserving branching assumption. We conclude that for realistic situations (the volume-preserving branching and the Kleiber law), the model predicts non-ergodic behavior.

Several measures of deviation from the ergodic behavior have been suggested in literature. Földes-Papp and Baumann (2011) suggested decoupling the effects of the molecular crowding and the temporal heterogeneity by presenting the power exponent, which controls the dynamics of the interaction network, as a product of these two factors.

Scott et al. (2009) suggested the ergodicity defect *D*, defined at different scales (on a map *T*) with respect to a basis of functions *f* given by an integral of the square of the space and time averages $D(f,T) \propto \int (f^*(x,T) - \bar{f})^2 dx$, where $f^*$ and $\bar{f}$ are the time and space averages. The ergodicity defect defined in this manner can be used to identify the structure of the solution, such as Lagrangian coherent structures (Rypina et al. 2011).

Using Eq. 10, one can suggest defining the ergodicity defect for our case as

$$D = (\theta - \rho)^2 = \rho^2 \left( \rho^{\frac{-2b-1+2c}{3b}} - 1 \right)^2 \tag{12}$$

Note that Eq. 12 does not involve the integration over space (i.e., L2 norm) because the parameters $\theta$ and $\rho$, as obtained from Eq. 10, do not depend on spatial variables. **Fig. 4(b)** shows a plot of the ergodicity defect vs. the volume fraction for the values of the parameters used in Fig. 4a.

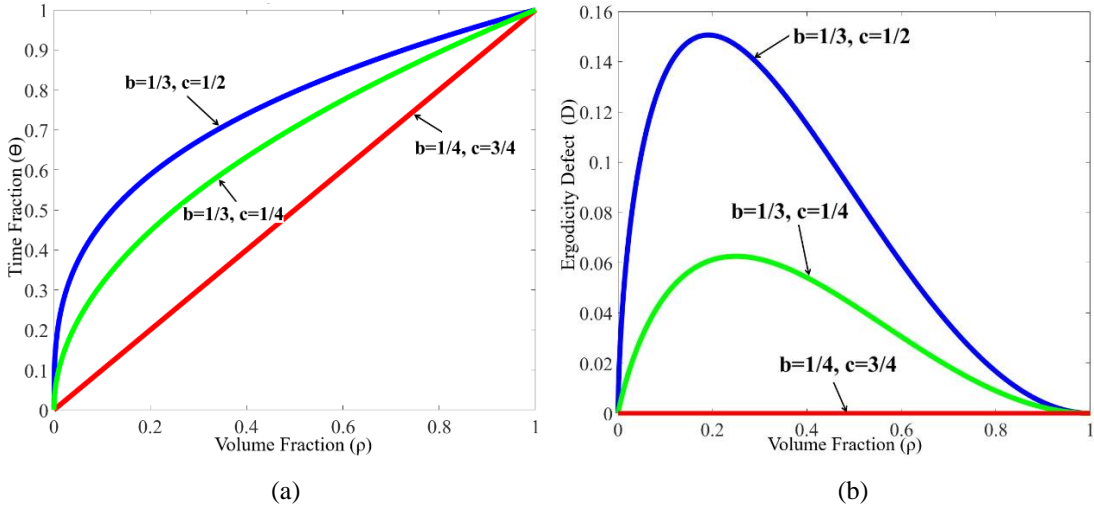

(a)                  (b)

Fig. 4. (a) Time fraction and (b) ergodicity defect as a function of the volume fraction for three combinations of parameters *a* and *b*.

## 4. Discussion

The calculations in the preceding section demonstrate that flow systems with fractal branching exhibit non-ergodic behavior. Such systems, including vascular networks, involve the flow, which covers a 3D volume delivering fluid to the vicinity of every point of the volume (every cell) with equal probability. In addition to covering the volume with equal probability, they should supply the fluid continuously in the temporal domain. However, due to the fractal organization of the vascular network, the flow "decelerates" at small length scales. Consequently, particles (including molecules and cells) spend more time at small volumes, in comparison with the volumetric fraction of these volumes. We interpret this feature of fractal behavior as ergodicity breaking.

While traditional sources of ergodicity breaking in biophysical fluid transport systems include time evolution or aging of the particles and macromolecular crowding, it is not uncommon that fractal geometry of a trajectory results in an effective deceleration of motion. One example would be the "hydrodynamic memory," which slows down the diffusion (Hofling and Franosch 2013). Another example is the so called "dissipative anomaly" in the turbulent flow, when the dissipation does not approach zero even at the zero viscosity limit, so that fractal trajectories lead to deceleration (De Lellis and Székelyhidi 2019; Shnirelman 2000).

Ergodicity breaking due to scale-dependent behavior is also significant for small systems of colloidal or droplet clusters (Nosonovsky and Roy 2020). These systems are used for *in situ* tracking of biomolecules and bioaerosols (Fedorets et al. 2019a). Lim et al. (2019) reported the transitions from sticky to ergodic configurations in six-particle and seven-particle systems of hard spheres. Fedorets et al. (2019b) studied small oscillations of a microdroplet cluster levitating in an ascending vapor stream and found that the cluster tends to oscillate as a whole. They showed that the synchronization of droplets' trajectories is not caused by the interactions between droplets, but by external fluctuations with the characteristic length scale much larger than the size of the droplets. The center of mass of the entire cluster produces the same motion as single micro-droplets, indicating ergodic behavior. Ergodicity breaking in this case is related to the scale ratio of the external fluctuation and of the system.

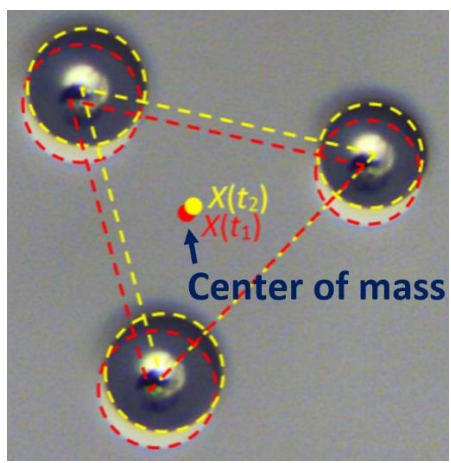

Figure 5. Oscillations of a small levitating droplet cluster, which vibrates as a whole (Fedorets et al. 2019b)

Understanding the ergodicity breaking in fluid flow of branching vascular networks is also relevant to micro/nanofluidic applications, such as the artificial biomimetic vascularized tissues. Such tissues are being developed for both medical applications of tissue engineered constructs (Hasan et al. 2014) and for applications such as self-healing materials where a flow of liquid agent is required (Nosonovsky and Rohatgi 2012).

**5. Conclusions**

Fractal models of branching cardiovascular networks involve two elements: the ability of a network to serve homogeneously a 3D volume (the volume preservation) and the conservation of the flow through the 2D cross-sections after branching (the area preservation). These models explain empirical allometric laws, such as the power exponent of ¾ in the Kleiber law, which relates metabolic rates to the animal mass. At the same time, these models raise new concerns, such as the need for more accurate accountability for capillary phenomena and more realistic dependencies for the blood flow velocity with decreasing size of vessels. We suggested a generalized formulation of the branching model (Eq. 3) and investigated the ergodicity of the fluid flow in a vascular network described by such a model. Generally, the fractal structure of such models makes them non-ergodic because a particle of the fluid spends more time in small 3D regions comparing with the volumetric fraction of these regions. The time fraction corresponds to the temporal average, while the volumetric fraction corresponds to the ensemble average.

The mechanism studied in the present paper can be a factor which contributes to the ergodicity breaking in biophysical systems, as follows from Eq. 10, in addition to such well-established mechanisms as aging and macromolecular crowding. Accounting for the non-ergodicity is important for a wide range of biomedical applications where long observations of time series are impractical.

**Acknowledgement**

Partially supported by the Russian Science Foundation (project 19-19-00076). The authors would like to thank Prof. Roshan D'Souza for the CFD software used in this study and anonymous reviewers for the discussion which improved this paper.